\documentclass[11pt]{article}
\usepackage{geometry}
\usepackage{a4}
\usepackage{graphicx}
\usepackage{epsf}
\usepackage{amsmath}
\usepackage{amssymb}
\usepackage{braket}
\usepackage{cite}
\newcommand{\be}{\begin{equation}}
\newcommand{\ee}{\end{equation}}

\newcommand{\Rmnum}[1]{\expandafter\@slowromancap\romannumeral #1@}
\newcommand{\bea}{\begin{eqnarray}}
\newcommand{\eea}{\end{eqnarray}}

\begin{document}
%%%%%%%%%%%%%%%%%%%%%%%%%%%%%%%
\def\A{{\mathbb{A}}}
\def\B{{\mathbb{B}}}
\def\C{{\mathbb{C}}}
\def\R{{\mathbb{R}}}
\def\s{{\mathbb{S}}}
\def\T{{\mathbb{T}}}
\def\Z{{\mathbb{Z}}}
\def\W{{\mathbb{W}}}
%%%%%%%%%%%%%%%%%%%%%%%%%%%%
\begin{titlepage}
\title{The geometry of RN-AdS fluids}
\author{}
\date{% authors are dated
Joy Das Bairagya, Kunal Pal, Kuntal Pal, Tapobrata Sarkar
\thanks{\noindent E-mail:~ joydas, kunalpal, kuntal, tapo @iitk.ac.in}
\vskip0.4cm
{\sl Department of Physics, \\
Indian Institute of Technology,\\
Kanpur 208016, \\
India}}
\maketitle
\abstract{
\noindent
We establish the parameter space geometry of a fluid system characterized by two constants, whose equation of state mimics that of the 
RN-AdS black hole. We call this the RN-AdS fluid. We study the scalar curvature on the parameter space of this
system, and show its equivalence with the RN-AdS black hole, in the limit of vanishing specific heat at constant volume. 
Further, an analytical construction of the Widom line is 
established. We also numerically study the behavior of geodesics on the parameter space of the fluid, 
and find a geometric scaling relation near its second order critical point.}
\end{titlepage}

\section{Introduction}

Black holes are singular solutions of general relativity that may arise as the end stages of gravitational collapse. 
Understanding the physics of black holes continues to be the focus of much attention. 
With the nature of black hole microstates being elusive, a popular line of research is to understand the 
macroscopic properties of black holes, for example their
thermodynamic properties, as these often provide useful insights into the underlying coarse grained structure of black holes. 
More recently, geometric methods have been applied to such studies, and it has been claimed in the literature 
that the nature of interactions between black hole microstates can possibly be elucidated by these analyses. 

The four laws of black hole mechanics which were formulated in the early 70's, are formally identical with the 
four laws of thermodynamics. For black holes having electric charge $Q$ and angular momentum 
$J$ with the corresponding potentials being $\Phi$ and $\Omega$ respectively, these are given 
(in geometric units) as :
{\bf (i)} The surface gravity $\kappa$ on the horizon of a stationary black hole of area $A$ is a constant.
{\bf (ii)} The change in energy $E$ of a black hole can be expressed as $dE = \kappa dA/8\pi + \Phi dQ + \Omega dJ$.
{\bf (iii)} The horizon area can never decrease with time, i.e $dA/dt \geq 0$.
{\bf (iv)} It is impossible to have $\kappa = 0$.
Hawking formalized this by showing that black holes can radiate, and that such radiation will have a temperature
$T=\kappa/2\pi$. Then, the four laws above are precisely the laws of thermodynamics, with the entropy
of the black hole identified with $A/4$. This formal correspondence has been studied extremely well,
for almost five decades now. 

Note that a pressure term is missing in the above discussion. For a long time it was believed that a pressure 
(and a volume) cannot be associated to black hole thermodynamics. This however changed a few  years back
with the proposal of Kastor, Ray and Traschen \cite{Ray} that a (varying) cosmological constant can be identified 
with the pressure of a black hole, with the conjugate volume associated with the volume of the event horizon. In this
formalism, the black hole mass has to be identified with the enthalpy of the system, and this can be shown to 
satisfy the corresponding Smarr relation. 
This is popularly termed as the extended phase space formalism in black hole thermodynamics.
Charged black holes
appearing in theories of gravity with a (negative) cosmological constant i.e the Reissner-Nordstrom-anti-de-Sitter (RN-AdS) 
black holes, where this is primarily interesting, will be the focus of this work. Thermodynamics of RN-AdS black holes
and their resemblance with van der Waals (vdW) systems were analyzed in the pioneering works 
of Chamblin, Emparan, Johnson and Myers \cite{Chamblin1},\cite{Chamblin2}. 
In these works, an equation of state that gave the temperature as a function of the charge and the electric potential,
was used to study this behavior. In the context of thermodynamics of the extended phase space, the results of \cite{Ray}  
were used by Kubiznak and Mann \cite{KubiznakMann} to derive a relation between the
pressure $P_{BH}$ and its volume ${\mathcal V}$ of the RN-AdS black hole of charge $Q$ and horizon radius $r_+$, 
via its Hawking temperature $T_{BH}$. This reads
\begin{equation}
P_{BH} = \frac{k_BT_{BH}}{2l_p^2r_+} - \frac{\hbar c}{8\pi l_p^2r_+^2} + \frac{\hbar c Q^2}{8\pi l_p^2r_+^4}~,~~
r_+ = \left(\frac{3{\mathcal V}}{4\pi}\right)^{\frac{1}{3}}~,
\label{original}
\end{equation}
where $k_B$ is the Boltzmann's constant, $\hbar$ is Planck's constant, $l_p$ is the Planck length, 
and $c$ is the speed of light. Upon restoring dimensions, $Q^2\equiv 
Q^2G/(4\pi\epsilon_0 c^4)$, with $G$ being Newton's constant and $\epsilon_0$ the permittivity of vacuum, 
carries dimension of length squared.
A large number of papers subsequently appeared that studied such relations in a variety of examples. 
On the other hand, the vdW equation of state of a fluid with pressure $P$ is given by eq.(76.7) of \cite{LL} as
\begin{equation}
P = \frac{Nk_BT}{V_s-Nb} - \frac{N^2a}{V_s^2}~, %+ \frac{d}{V_s^4}~,
\label{vdw}  
\end{equation}
where $N$ denotes the number of particles (molecules), and $V_s$ the volume of the fluid. 
Now using $V = V_s/N$ which denotes the volume per molecule (so that the molecular density is $\rho = 1/V$), 
and introducing a further $V_s^{-4}$ dependence in the virial expansion (see section $75$ of \cite{LL}), 
we can write a modified form of eq.(\ref{vdw}) that is formally identical to eq.(\ref{original}) with $b=0$, and given by 
\begin{equation}
P = \frac{k_BT}{V} - \frac{a}{V^2} + \frac{d}{V^4}~,
\label{vdwmod}  
\end{equation}
with $P=P_{BH}$, $T=T_{BH}$, if we identify $V=2l_p^2r_+$, 
and in addition identify the constants $a = \hbar c l_p^2/(2\pi)$ and $d=2\hbar cQ^2l_p^6/\pi$ which 
carry dimensions of energy times volume, and energy time volume cubed, respectively. If such an identification
is not done, and we allow for generic values of $a$ and $d$, then eq.(\ref{vdwmod}) represents the equation
of state of a fluid that we will call the RN-AdS fluid, with the RN-AdS black hole being the special case
where the above identification is imposed. 
In the original vdW equation of state $P_{vdW}=k_BT/(V-b) - a/V^2$, the constant $b$ takes care of the
fact that the volume per molecule cannot be lower than a certain finite hard-sphere cutoff, related to $b$. Although it 
might seem that this condition is relaxed here, we will see later that there is indeed such a minimum
volume allowed by the system, which is pressure dependent. The physical reason for this is clear : the last
term in eq.(\ref{vdwmod}) is a repulsive term, due to which this minimum volume arises, and it will go to
zero only for infinite pressure. In this paper, we will treat eq.(\ref{vdwmod}) as a phenomenological equation of state,
without delving into a statistical derivation, aspects of which will be discussed elsewhere. 

It is known for some time that the vdW system (and for that matter any system in thermodynamic equilibrium) is 
amenable to a geometric analysis. Such analyses began with the work of Ruppeiner \cite{Rupp}, was extended to the case of 
quantum systems by Provost and Vallee \cite{PV}, and is by now an established tool for the study of phase transitions 
in classical and quantum systems. The geometry (also called information geometry in the literature) 
is that of the parameter manifold, with coordinates being the 
tunable parameters of the theory such as the temperature and the density, on which one can define a
Riemannian metric. The Ricci scalar curvature $R$ of this metric has interesting properties, namely, it diverges along the spinodal 
curve, and is conjectured to be related to the correlation volume, i.e scales as the correlation volume near criticality.
This conjecture was proved
via a renormalization group analysis, in \cite{tapo5}. Further, with the above identification
of $R$, its equality in the liquid and gas phase was used to predict first order phase transitions. Also, the extrema 
of $R$ can be computed analytically, which provides a tool to compute the Widom line \cite{Widom}, \cite{Widom1},
which is an extension of the phase coexistence line beyond criticality and is defined as the locus of the
extrema of the correlation length in this region. 
In \cite{tapo1}, all these facts about the geometry of the vdW fluid was compared with experimental data from 
the NIST database \cite{NIST}, and excellent agreement was found in cases of simple fluids that closely follow the vdW equation 
of state, both in the sub-critical and the super-critical regions. 

In this work, we establish the geometric framework for studying a fluid, which we have called the RN-AdS fluid, that satisfies 
the equation of state of eq.(\ref{vdw}) with generic values of $a$ and $d$
(the geometry of RN-AdS black holes without the extended phase space notion
was worked out in \cite{tapo1b}, see also \cite{Sengupta}). 
Our RN-AdS fluid may have a non-zero specific heat at constant volume $c_v$ (we choose
$c_v=3/2$ for illustrations here) or, as a limiting case, it might have $c_v \to 0$. In this latter case, following the
work of \cite{WeiLiuMannPRL}, we contrast our results with the RN-AdS black hole of eq.(\ref{original}) 
for which an analysis of the Helmholtz free energy shows that $c_v$ vanishes identically. 
The scalar curvature on the parameter manifold 
of the RN-AdS fluid with $c_v\to 0$ is shown to be the same as the RN-AdS black hole, and we thus argue why
it might be possible to glean insight into the extended phase space thermodynamics of such black holes
via the RN-AdS fluids, rather than vdW systems. 
We further study geodesics on the parameter manifold arising in our fluid system, and compute a geometric scaling relation 
near criticality. These are shown to be qualitatively similar to the vdW fluid and we also comment on the case where
$c_v$ vanishes, which offers a ready comparison with the the RN-AdS black hole. 
The scalar curvature gives us an analytical handle to compute the
Widom line, and we show that this has a different nature for the case $c_v \neq 0$, compared to the case
of vanishing $c_v$. 

This paper is organized as follows. In the next section \ref{sec2}, we briefly recall Ruppeiner's geometric formalism,
and point out some physical features that have to be respected in such analyses. Next, in section \ref{sec3}, we
establish the geometry of RN-AdS fluids and comment on their relation to the RN-AdS black hole. The Widom line
is also computed here. Section \ref{sec4}
studies geodesics on the parameter manifold of RN-AdS fluids and the corresponding black holes. Next, in
section \ref{sec5}, we define and compute a geometric critical exponent for the fluid system. Section \ref{sec6}
ends with some discussions on our main results.

\section{Ruppeiner's Geometry}
\label{sec2}

Ruppeiner's formulation of the geometry of the parameter manifold of thermodynamic and statistical systems
starts with assigning the system a fixed volume $V_s$, which is in equilibrium with a reservoir of volume $V_c$. 
The entropy of such a statistical system is given by Boltzmann's formula,
\begin{equation}
S=k_B \log\Omega~,
\end{equation}
where $\Omega$ is the number of the microscopic states of the corresponding thermodynamic system 
and $k_B $ is the Boltzmann constant. For a system of $n$ independent variables (i.e tuning parameters) 
$x^\alpha, \alpha =1,2, \cdots n$, the probability of finding its state
between $(x^1 , \cdots, x ^n )$ and $(x ^1 + dx^ 1 , \cdots, x ^n + dx^ n )$ is proportional to the number of microstates 
\begin{equation}
P(x^1,....,x^n)=C\Omega(x^1,....,x^n)dx^1dx^2....dx^n~,
\end{equation}
where $C$ is a normalization constant. Hence, one has 
\begin{equation}
P(x^1,...,x^n)\propto e^ \frac{S}{k_B}~.
\end{equation}
Expanding $S$ about a local equilibrium value $S_0$, one obtains in the limit $V_c \gg V_s$ \cite{Rupp},
\begin{equation}
P\propto e^{-\frac{V_s}{2}\Delta l^2}~,~~{\rm with}~,~~
\Delta l^2 = -\frac{1}{V_sk_B} \frac{\partial ^2 S}{\partial x^\alpha \partial x^ \beta}\Delta x^\alpha \Delta x^\beta ~.
\label{dimprob}
\end{equation}
In thermodynamic geometry, $\Delta l^2$ measures the distance between two neighbouring 
states that are related by a fluctuation, and the metric on the space of parameters is defined from 
\begin{equation}
\Delta l ^2 = g_{\alpha _\beta} \Delta x^\alpha \Delta  x^\beta~,~~g_{\alpha\beta}
=-\frac{1}{V_sk_B} \frac{\partial ^2 S}{\partial x^\alpha \partial x^ \beta}~.
\label{metelements}
\end{equation}
Taking this as the line element, one can compute the Ricci scalar curvature $R$. From now, we specialize to
the case $n=2$. If the metric is diagonal, then the scalar curvature for the metric can then be obtained from
the formula 
\begin{equation}
R = \frac{1}{\sqrt{g}}\left[\frac{\partial}{\partial x^1}\left(\frac{1}{\sqrt{g}} \frac{\partial g_{22}}{\partial x^1}\right) + 
\frac{\partial}{\partial x^2}\left(\frac{1}{\sqrt{g}} \frac{\partial g_{11}}{\partial x^2}\right)\right]~,~~g=\sqrt{g_{11}g_{22}}~,
\label{scalarcurvature}
\end{equation}
with a slightly more complicated expression for $R$ for non-diagonal metrics. 
The fact that $R$ is related to a correlation length was used
in \cite{tapo1} to predict first order phase transitions in vdW systems, via the equality of $R$ in the
liquid and gas phases. This gave a geometric way of computing the phase co-existence line, 
and bypassed a few traditional problems with the Maxwell construction, and was shown to 
be in good agreement with experimental data. 

There are several equivalent ways to write the metric of eq.(\ref{metelements}). A  popular representation
involves the Helmholtz free energy per unit volume. In this representation, 
one uses the notion of the specific entropy $s= S/V_s$ for a {\rm fixed} volume $V_s$ of the system
with $N$ particles (for a pedagogical exposition, see \cite{RuppAJP}). 
Now, if we consider the quantities $T$ and $\rho = N/V_s=1/V$ as the fluctuating parameters, 
with $V_s$ held fixed (so that $N$ fluctuates), then the line element is conveniently written as 
\begin{equation}
dl^2 = \frac{1}{k_B T}\left(\frac{\partial s}{\partial T}\right)_{\rho} dT^2+ \frac{1}{k_B T}
\left(\frac{\partial \mu}{\partial \rho}\right)_Td\rho^2
\label{line}
\end{equation}
where $\mu = \left(\frac{\partial f}{\partial \rho}\right)_T$, $f$ being the Helmholtz free energy per unit
volume. This definition of the line element yields a scalar curvature that has the dimension of volume,
and can be meaningfully compared to the correlation volume near criticality. 
It is well knows that for real fluid systems, the scalar curvature $R$ satisfies the following properties : 
{\bf (a)} It has the dimension of volume, with the limit of applicability of the geometric formalism being restricted to 
cases where $R$ is greater than a few molecular volumes \cite{tapo1}. 
{\bf (b)} For systems that exhibit first and second order phase transitions, it diverges 
along the spinodal curve where $(\partial P/\partial V)=0$. 
{\bf (c)} It diverges at criticality with a power law exponent $t^{-2}$ with $t$ denoting the deviation from the 
reduced critical temperature. 

We recall and emphasize here that Ruppeiner's formalism assumes at the outset that there is a system 
of fixed volume $V_s$ and a surrounding of volume $V_c \gg V_s$, and that the entropy of the system and the surrounding scale 
with the volumes in a similar fashion \cite{Rupp}. This is of course difficult to envisage for black hole 
systems where the black hole entropy scales as an area, 
and the geometric formalism is more challenging here. In the absence of a notion of black hole volume
(i.e in the pre extended phase space era), several authors thus used the definition of the line element
that follows directly from eq.(\ref{dimprob}), i.e 
\begin{equation}
g_{\alpha\beta} = -\frac{1}{k_B}\frac{\partial^2 S}{\partial x^{\alpha}\partial x^{\beta}}
\label{dimprob1}
\end{equation}
Although it still gives the features {\bf (b)} and {\bf (c)} discussed in the previous paragraph, as checked
in several black hole examples till date,
it does not satisfy condition {\bf (a)} there. This is because if we write the line element in the form of 
eq.(\ref{dimprob1}), then the quantity $\Delta l ^2$ is dimensionless, and so is the scalar curvature $R$, 
as can be seen from eq.(\ref{scalarcurvature}). It is therefore difficult to assign a physical meaning to $R$ 
(as a correlation volume), if it is computed from eq.(\ref{dimprob1}). As we have mentioned, in the sub-critical
region of a vdW system, the analysis of \cite{tapo1} showed that the first order phase transitions can
be determined from an $R$-matching method, which relies on the fact that the correlation lengths of 
the liquid and the gas phases become equal at such a transition.
This analysis will also not carry much meaning if $R$ is dimensionless. 

However, in eq.(\ref{original}), if we interpret the black hole volume ${\mathcal V}$ as the volume per ``microstate'' of the system, 
then it is possible to use the temperature and density as coordinates, similar to the case of RN-AdS fluids, 
and draw physical conclusions out of a geometric analysis. In this case, one has to simply use the second relation 
of eq.(\ref{metelements}) as the appropriate definition of the metric.
Of course, the mathematical difference between values of $R$ computed from eq.(\ref{dimprob1}) as compared
to that computed from the second relation of eq.(\ref{metelements}) is easy to guess. Since we are treating
volume as fixed, the expression for $R$ obtained from the first should simply be the system volume 
times the one obtained from the second. As we will see, this is indeed true. 

\section{Geometry of RN-AdS Fluids}
\label{sec3}

With the discussion above, we are now ready to present our main computation. For the RN-AdS fluid, 
we start with the ideal
gas, whose free energy per unit volume in terms of its temperature $T$ 
follows from the standard textbook definition from Landau and Lifshitz \cite{LL}
\begin{equation}
f_{id} = -\rho k_B T\log(e/\rho) + \rho h(T)~,
\label{fidLL}
\end{equation}
where we recall that $\rho = N/V_s=1/V$, where $V_s$ is held fixed and $N$ denoting the number of molecules. 
here $h(T)$ is a function of $T$ from which we can define $k_Bc_v=-Th''(T)$, so that we can write
$h(T) = -c_vk_BT\log(T/e)$. Here, $c_v$ is the dimensionless specific heat at constant volume, per molecule. 
By a standard procedure analogous to the van der Waals case, we write the free energy per unit volume of the
RN-AdS fluid as 
\begin{equation}
f(T,\rho) = f_{id} - \rho k_BT\log(1-b\rho) - a\rho^2 + \frac{d}{3}\rho^4~,
\label{freeenergy}
\end{equation}
where we will set $b=0$, following the discussion in the introduction. The vdW case corresponds to a non-zero value 
of $b$ with $d=0$. For the RN-AdS fluid, we then have from eqs.(\ref{fidLL}) and (\ref{freeenergy}), 
the free energy per unit volume given as, 
\begin{equation}
f(T,\rho)= -\rho k_BT\log(e/\rho) - \rho c_vk_BT\log(T/e) - a\rho^2+\frac{d}{3}\rho^4~.
\label{FEfull}
\end{equation}
The pressure follows from the standard expression $P = -f + \rho(\partial f/\partial \rho)$ and gives the
same expression as eq.(\ref{vdw}) after converting to the variable $V$. 
By plotting isotherms (which we show in sequel) or the free energy, it can be seen that this model gives the usual 
first order liquid-gas phase transition culminating in a second order critical point. 

Now, we note that there is a minimum volume $V_{min}$ dictated by eq.(\ref{vdwmod}) so that the temperature does not become negative.
This reads,
\begin{equation}
V_{min} = \frac{1}{\sqrt{2P}}\left[\left(a^2+4Pd\right)^{1/2} - a\right]^{1/2}
\label{volmin}
\end{equation}
In the limit $P \to \infty$, this goes to zero, but has a finite value $V_{min} = \sqrt{d/a}$ at $P=0$.
This is interesting, as for the vdW equation of state, the minimum volume required to 
prevent the temperature from being negative is $V_{min, vdW}=b$, as mentioned in the introduction. 
So although we have set $b=0$ for the RN-AdS fluid, the notion of a minimum volume is inbuilt in
its construction. The interpretation of this is as mentioned in the introduction. Namely, the
notion of a minimum volume occurs due to the repulsive nature of the last term in eq.(\ref{vdwmod}). 
Whereas the vdW equation of state has a hard sphere cut-off as the minimum volume, here we 
recover such a notion from the repulsion between the molecules. 
This will have important consequences as we illustrate in the discussion to follow. 
We will remember that from eq.(\ref{volmin}), it follows that as $P\to \infty$, $V_{min}\sim P^{-1/4}$. 

Next, we recall that the critical point is a point of inflexion, determined from
\begin{equation}
\left(\frac{\partial P}{\partial V}\right)_T = \left(\frac{\partial^2 P}{\partial V^2}\right)_T=0~.
\end{equation}
Using eq.(\ref{vdw}), we find that the solution to the above equations give
\begin{equation}
T_c = \frac{2\sqrt{2} a^{\frac{3}{2}}}{3\sqrt{3}k_B\sqrt{d}}~,~~V_c = \frac{\sqrt{6d}}{\sqrt{a}}~,~P_c = \frac{a^2}{12d}~,
\label{tcpcvc}
\end{equation}
where the last relation follows by putting the values of $T_c$ and $V_c$ in eq.(\ref{vdw}). 
Using $T_r = T/T_c$, $V_r = V/V_c$ and $P_r = P/P_c$, we can now write the analogue of the VdW equation
of eq.(\ref{vdw}) as
\begin{equation}
P_r = \frac{8T_r}{3V_r} - \frac{2}{V_r^2} + \frac{1}{3V_r^4}~.
\label{red}
\end{equation}
%%%%%%%%%%%%%%%%%%%%%%%%%%%%%%%%%%%%%%%%%%%%%%%%%%
\begin{figure}[h!]
\centering
\includegraphics[width=3in,height=2in]{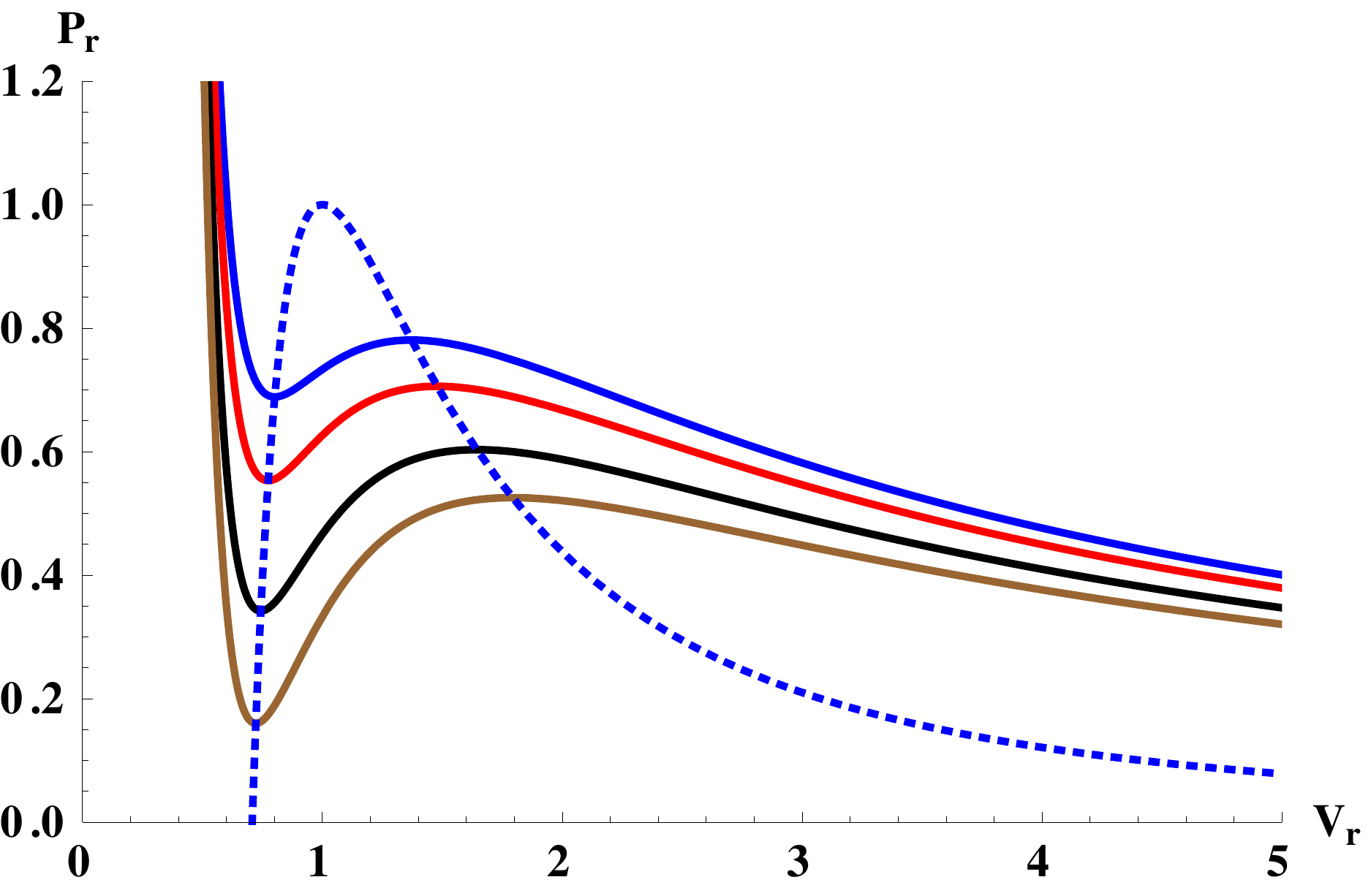}
\caption{Isotherms and spinodal curve.}
\label{fig1}
\end{figure}
%%%%%%%%%%%%%%%%%%%%%%%%%%%%%%%%%%%%%%%%%%%%%%%%%%
It is not difficult to compute the metric components of eq.(\ref{line}). We do this in the $T,\rho$ coordinates,
and obtain in terms of $T$ and $\rho$,
\begin{equation}
g_{TT} = \frac{\rho c_v}{T^2}~,~~g_{\rho\rho}=\frac{k_BT - 2a\rho + 4d\rho^3}{k_BT\rho}
\label{mettrho}
\end{equation}
Using the value $c_v=3/2$ for illustration, we obtain the value of the scalar curvature which is
conveniently expressed in terms of the reduced variables as 
\begin{equation}
R = \sqrt{\frac{2d}{3a}}\frac{V_r \left[1-V_r^2\left(6-9V_r^2+T_rV_r\left(5+3V_r^2\right)\right)\right]}{\left(V_r^2 (2 T_r
V_r-3)+1\right)^2}
\label{gencvR}
\end{equation}
For RN-AdS black holes which we will come to in a moment, 
the limit $c_v \to 0$ is important, and here we obtain (using the notation of \cite{WeiLiuMannPRL}) for
the RN-AdS fluid,
\begin{equation}
R_{n}=Rc_v = -\sqrt{\frac{3d}{2a}}\frac{V_r\left(3 V_r^2-1\right)\left(4 T_r V_r^3 - 3V_r^{2}+
1\right)}{\left(2 T_r V_r^3-3 V_r^{2}+1\right)^2}= -\frac{V_cV_r\left(3 V_r^2-1\right)\left(4 T_r V_r^3 - 3V_r^{2}+
1\right)}{2 \left(2 T_r V_r^3-3 V_r^{2}+1\right)^2}
\label{cvzeroR}
\end{equation}

From eq.(\ref{gencvR}) or the first relation of eq.(\ref{cvzeroR}), since the constant $a$ carries the dimension 
of energy times volume and $d$ has dimension energy times volume cubed, clearly the dimension of $R$ is that of volume. 
This information is carried by $V_c$ in the second relation of eq.(\ref{cvzeroR}). The first condition listed at the end of 
the last section is thus satisfied. Now, it is seen that $R$ diverges along the curve given 
by $T_s = (3V_r^2 - 1)/(2V_r^3)$.
In fig.(\ref{fig1}), we plot the isotherms corresponding to $T_r = 0.9$ (solid blue), $0.86$ (solid red), 
$0.80$ (solid black) and $0.75$ (solid brown). The dotted blue line corresponds to the 
$V_r - P_r$ curve with $T = T_s$. It is clearly seen that this is the spinodal curve. The second condition
of the last section is thus satisfied as well. Also, since $V_cV_r=V$, we can write eq.(\ref{cvzeroR}) more
conveniently as a curvature per unit volume, 
\begin{equation}
{\mathcal R} = \frac{Rc_v}{V} = -\frac{\left(3 V_r^2-1\right)\left(4 T_r V_r^3 - 3V_r^{2}+
1\right)}{2 \left(2 T_r V_r^3-3 V_r^{2}+1\right)^2}~.
\label{puv}
\end{equation}
Importantly, we see that if we substitute the reduced temperature in terms of the reduced 
pressure in eq.(\ref{red}), the we obtain from eq.(\ref{puv}) that in terms of $P_r$ and $V_r$, we have
\begin{equation}
{\mathcal R} = -\frac{4}{9}\frac{\left(3 V_r^2-1\right)\left(1+3P_rV_r^4\right)}{\left(1-2V_r^2+P_rV_r^4\right)^2}~.
\label{tv}
\end{equation}
In eq.(\ref{tv}) above, since $P_r$ and $V_r$ are used as independent variables, we can see that 
for a given non-zero value of $V_r$, ${\mathcal R} \to 0$ as $P_r\to \infty$. Importantly,  
$R, R_n \sim P_r^{-1}$, i.e in terms of the original variables, $R, R_n  \sim P^{-1}$. 
Recalling that $V_{min}\sim P^{-1/4}$ as discussed earlier, we see that
in the limit of high pressure, $R$ goes to zero more rapidly for finite values of $V_r$ compared
to $V_{min}$. This sets the limits of applicability of the theory, namely that in the regimes of
high pressure (or high temperatures with fixed small values of $V_r$ 
as follows from eq.(\ref{red})), the geometric analysis
breaks down. The correlation volume here becomes less than the minimum volume allowed by
the theory, and hence the Gaussian approximation on which this is based, becomes invalid. 
This result is consistent with the known fact that for vdW fluids, the geometrical picture
is strictly valid only for $P_r \leq 10$ in the supercritical regime \cite{tapo1}. 
%%%%%%%%%%%%%%%%%%%%%%%%%%%%%%%%%%%%%%%%%%%%%%%%%%
\begin{figure}[h!]
\centering
\includegraphics[width=3in,height=2in]{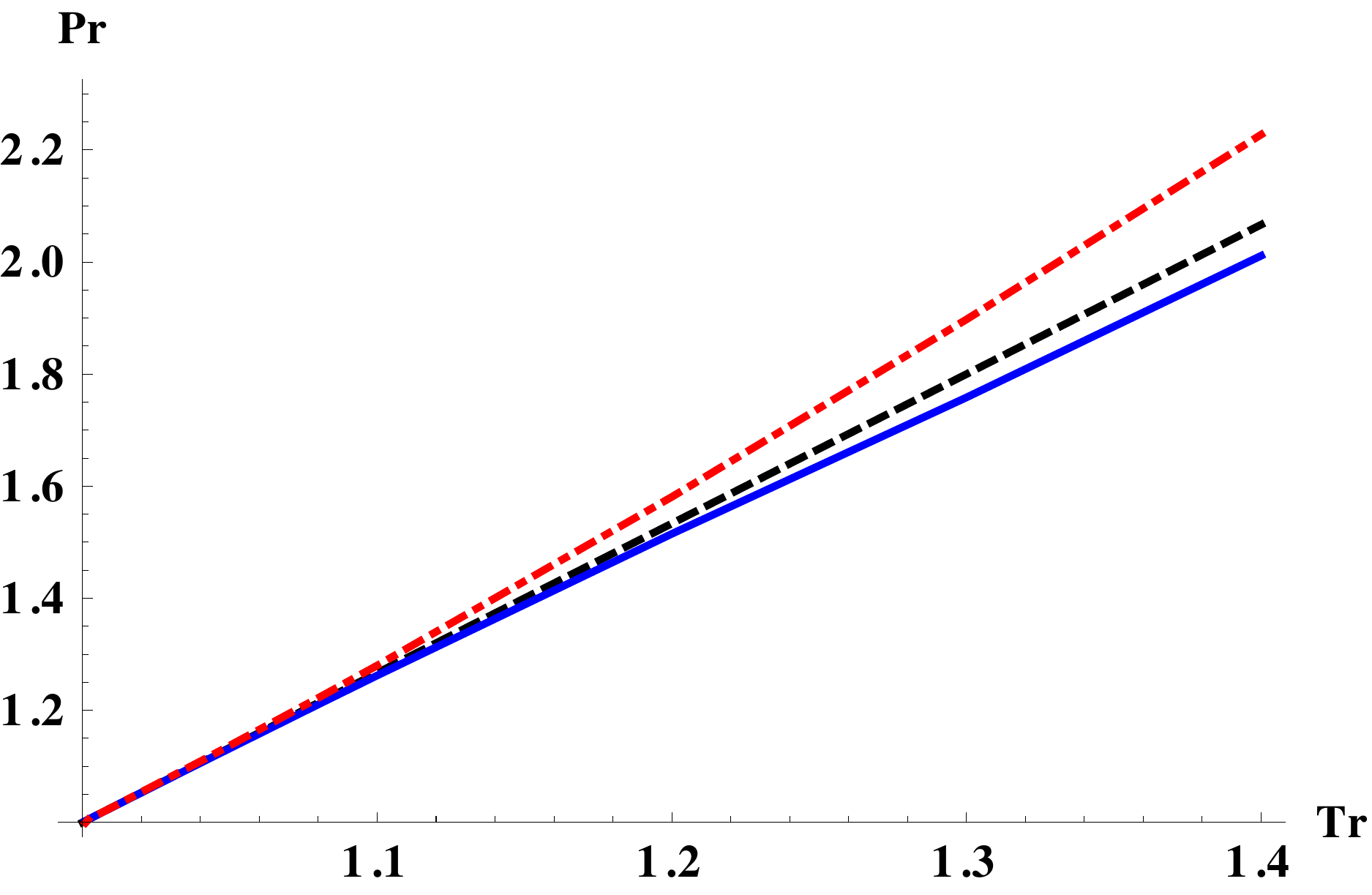}
\caption{Widom line with $c_v=3/2$ (dot-dashed red), $c_v=0$ (solid blue). The maxima of $c_P$
is shown in dashed black.}
\label{fig1b}
\end{figure}
%%%%%%%%%%%%%%%%%%%%%%%%%%%%%%%%%%%%%%%%%%%%%%%%%%

We now comment on the Widom line \cite{Widom}, \cite{Widom1}, which is an extension of the liquid-gas
coexistence curve in the supercritical region. This is analytically defined as the locus of maxima of the
correlation length. With the (magnitude of the) scalar curvature $R$ identified as the correlation volume, 
the Widom line in our case then is simply the locus of extrema of $R$. We plot this, from eqs.(\ref{gencvR})
and (\ref{cvzeroR}) for the cases $c_v=3/2$ and $c_v=0$, respectively, in fig.(\ref{fig1b}). These are shown 
by the dot-dashed red line ($c_v=3/2$) and the solid blue line ($c_v = 0$). Here, the specific heat
at constant pressure is given by 
\begin{equation}
c_p = c_v + \frac{3T_rV_r^3}{4\left(1-3V_r^2 + 2T_rV_r\right)}
\end{equation}
The locus of maxima of $c_p$ is then given by $V_r=1$, i.e $P_r= -5/3 + 8T_r/3$. This is plotted as
the dashed black line in fig.(\ref{fig1b}). We see that while close to criticality, the three lines are
indistinguishable, they differ away from criticality. This is a major difference between the RN-AdS fluid
with $c_v=3/2$ and the RN-AdS fluid with $c_v=0$. 

We now turn to the RN-AdS black hole. As a convenience, we will set $G=c=l_p=k_B=1$, since our final results
will only depend on reduced variables. The free energy in geometric units is given by \cite{KubiznakMann}
\begin{equation}
F = \frac{1}{2}\left(r_+ - 2\pi T r_+^2 + \frac{Q^2}{r_+}\right)~.
\label{FE}
\end{equation}
We begin with the free energy per unit volume. There are two distinct notions of volume here, as pointed out in
the introduction. Namely, one is $V=2l_p^2r_+$ and the other is ${\mathcal V} = (4/3)\pi r_+^3$. The latter
being the volume of the system (i.e the black hole), it is more appropriate to use 
\begin{equation}
\frac{F}{{\mathcal V}} = -\left(\frac{9\pi}{16}\right)^{1/3}T{\tilde\rho}^{1/3} 
+ \left(\frac{3}{32\pi}\right)^{1/3}{\tilde\rho}^{2/3}+ 
\left(\frac{\pi}{6}\right)^{1/3}Q^2{\tilde\rho}^{4/3}~,
\label{FE1}
\end{equation}
where we define the density per microstate, ${\tilde \rho} = 1/{\mathcal V}$. We now use the line element of eq.(\ref{line})
in $(T,{\tilde \rho})$ coordinates. 
Since $c_v=0$ here, following \cite{WeiLiuMannPRL}, we will assume the metric component 
$g_{TT}=c_v/T^2$ and then take the limit $c_v \to 0$. Finally, we obtain using
${\mathcal V}_c=8\sqrt{6}\pi Q^3$, $T_c=\sqrt{6}/(18\pi Q)$ and $P_c = 1/(96\pi Q^2)$,
\begin{equation}
R_n=Rc_v = -\frac{{\mathcal V}_c{\mathcal V}_r\left(3{\mathcal V}_r^{2/3}-1\right)\left(4T_r{\mathcal V}_r
-3{\mathcal V}_r^{2/3} +1\right)}{2\left(2T_r{\mathcal V}_r - 3{\mathcal V}_r^{2/3}+1\right)^2}
\label{properR}
\end{equation}
Apart from a factor of ${\mathcal V}_c{\mathcal V}_r={\mathcal V}$, 
this is the same result as obtained in \cite{WeiLiuMannPRL}. Now
we see that $Rc_v$ in eq.(\ref{properR}) carries the dimension of volume as it should, and it depends
on the particular RN-AdS black hole through ${\mathcal V}_c$, via the charge $Q$. Also note that for the
case $c_v=0$, the reduced volume ${\mathcal V}_r = V_r^3$ used in the RN-AdS black hole (see discussion in
the introduction). This follows if we write the quantities $a$ and $d$ for the RN-AdS fluid in terms of the 
RN-AdS black hole via their equations of state (or the values of the critical quantities), to obtain
\begin{equation}
a = \frac{1}{2\pi}~,~~d = \frac{2Q^2}{\pi}~.
\label{identi}
\end{equation}
We remind the reader that eq.(\ref{identi}) is valid only for the RN-AdS fluid with $c_v=0$. 
Using this, we can readily compare eq.(\ref{properR}) with eq.(\ref{puv}). 
Eq.(\ref{properR}) is seen to imply that 
\begin{equation}
{\mathcal R} = \frac{Rc_v}{{\mathcal V}}=-\frac{\left(3V_r^{2}-1\right)\left(4T_rV_r^3
-3V_r^{2} +1\right)}{2\left(2T_rV_r^3 - 3V_r^{2}+1\right)^2}
\label{puv1}
\end{equation}
Comparing eqs.(\ref{puv}) and (\ref{puv1}), we see that these are the same. This is important,
as we have effectively shown that the RN-AdS fluid with $c_v\to 0$ and the
RN-AdS black hole are described by the same scalar curvature per unit volume. Note that
the free energy per unit volume, from which these are calculated have different forms in this limit (compare
eqs.(\ref{FEfull}) and (\ref{FE}).

At this point, we recall the result for vdW fluids of eq.(\ref{vdw}) given in \cite{tapo1},
\begin{equation}
\frac{Rc_v}{V} = -\frac{\left(3V_r - 1\right)^2\left(8T_rV_r^3 - (1-3V_r)^2\right)}{2\left(4T_rV_r^3-
(1-3V_r)^2\right)^2}~,
\label{puvvdw}
\end{equation}
with $V=V_rV_c$, $V_c=3b$. We see that eq.(\ref{puvvdw}) is indeed quite different from
eq.(\ref{puv}) or eq.(\ref{puv1}). 
It therefore seems that a comparison of the geometry of the RN-AdS black hole with that of the
vdW system may be somewhat inappropriate. The correct comparison of the black hole should be 
with our RN-AdS fluid in the limit $c_v\to 0$.

\section{Geodesics on the parameter manifold for RN-AdS fluids}
\label{sec4}

\begin{figure}[h!]
\centering
\includegraphics[width=3in,height=2in]{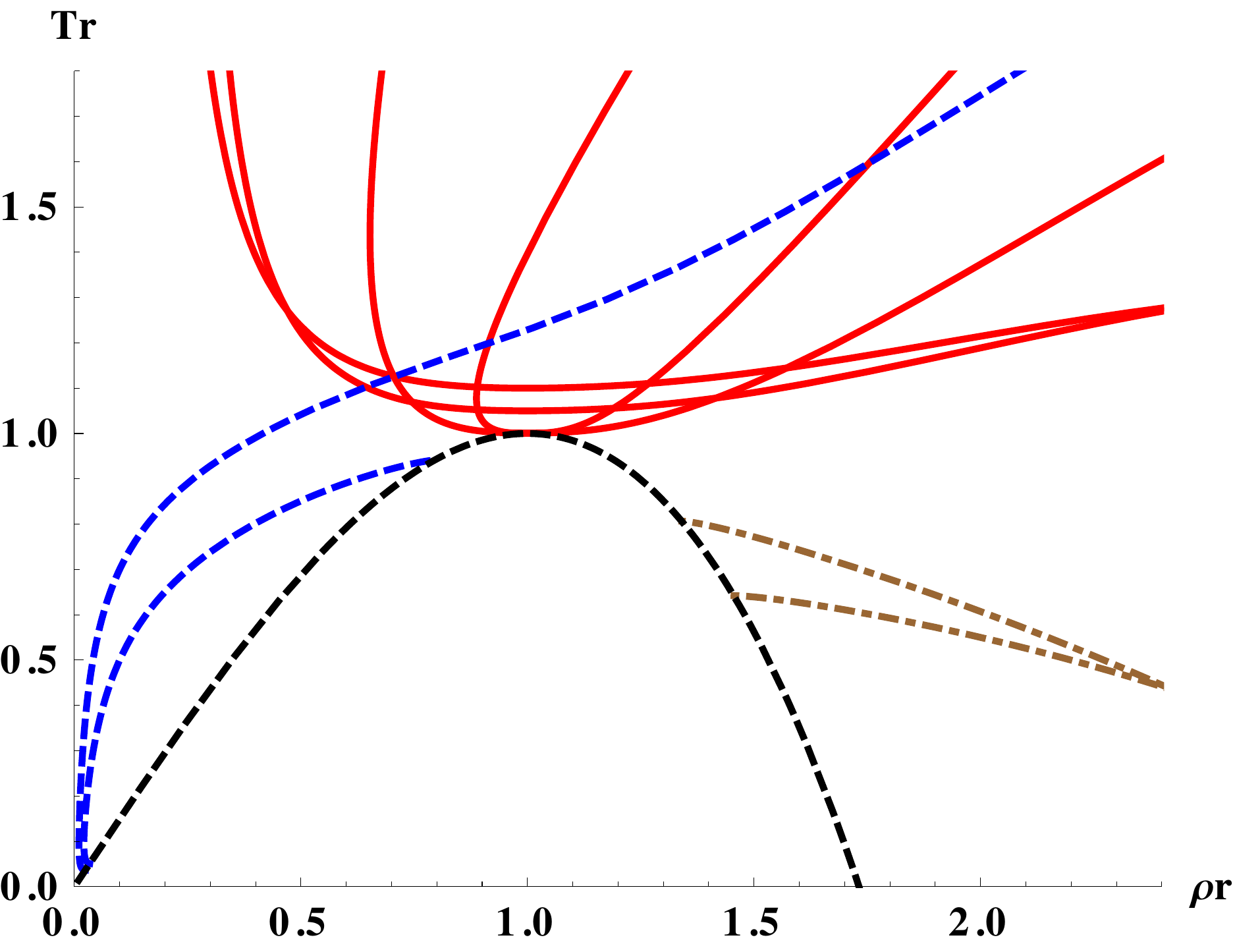}
\caption{Geodesics on the parameter manifold.}
\label{fig2}
\end{figure}

We will now comment on the geodesic equations on the parameter manifold for RN-AdS fluids. 
To perform explicit computations, it is convenient to set $T_c = V_c = 1$, in which case one obtains
from eq.(\ref{tcpcvc}) that $a = 3/4$ and $d = 1/8$. We will also set $k_B=1$ and take $c_v = 3/2$ for illustration. 
Using the metric of eq.(\ref{mettrho}), we then find that
the coupled non-linear geodesic equations in the $T,\rho$ representation are given by 
\begin{eqnarray}
&~&{\ddot T}+\frac{\left(\rho^2-3\right) {\dot \rho}^2}{6 \rho}+\frac{{\dot \rho}{\dot T}}{\rho}-\frac{{\dot T}^2}{T}=0\nonumber\\
&~&{\ddot \rho} - \frac{3\rho{\dot T}^2}{2T{\mathcal A}} - \frac{\left(\rho^3 - 3\rho\right){\dot T}{\dot \rho}}{T{\mathcal A}}
-\frac{\left(T-\rho^3\right){\dot \rho}^2}{\rho{\mathcal A}}=0~,
\label{geoeq}
\end{eqnarray}
where ${\mathcal A} = 2T - 3\rho +\rho^3$, and $T$ and $\rho$ are considered to be functions of an affine parameter
with respect to which the derivatives are taken. Now we use a standard numerical recipe to integrate the geodesic
equations along with specified boundary conditions. Namely, we shoot geodesics from specific points on the
manifold towards the spinodal curve, and numerically solve eq.(\ref{geoeq}). Our numerical results are 
shown as a plot of such several such geodesics in the $\rho-T$ plane 
in fig.(\ref{fig2}), where the spinodal curve is shown in dashed black. 
%%%%%%%%%%%%%%%%%%%%%%%%%%%%%%%%%%%%%%%%%%%%%%%%%%
%%%%%%%%%%%%%%%%%%%%%%%%%%%%%%%%%%%%%%%%%%%%%%%%%%
These show expected behavior in lines with the discussion in \cite{tapo6}. Namely, geodesics can
terminate on the spinodal curve, away from the second order phase transition. Near the singularity,
these typically show turning behavior. Also, geodesics are typically confined to one phase, i.e a geodesic
originating in the liquid phase will not cross over to the gas phase, and vice versa. In principle this analysis might be extended to the case of
the RN-AdS fluid with $c_v=0$. We have numerically checked that using a very small value
of $c_v$, the behavior of geodesics in that case is qualitatively similar to the one presented in
fig.(\ref{fig2}). Hence, the behavior of geodesics in the parameter space of RN-AdS black holes 
is similar to the RN-AdS fluid discussed here. In the next section, we will comment on the behavior of geodesics close to 
criticality. 

\section{A Geometric scaling relation for RN-AdS fluids}
\label{sec5}

In \cite{tapo4}, it was pointed out that there are important relations between scalar quantities on the parameter
manifold of classical and quantum systems, near criticality. In particular, it is natural to express scaling invariants
in terms such scalar quantities. Apart from the Ricci scalar curvature, on the two dimensional parameter
manifolds one such quantity is the expansion parameter of geodesics. Let us review this in brief. 

If $x^{\mu}$ generically denotes the coordinates on a curved manifold, then the geodesic equation is
${\ddot x^{\mu}} + \Gamma^{\mu}_{\nu\rho}{\dot x^{\nu}}{\dot x^{\rho}} = 0$, 
where the usual Christoffel connections are $\Gamma^{\mu}_{\nu\rho}$ 
and derivatives are taken with respect to an affine parameter along the geodesic. The tangent vector
to the geodesic is  $u^{\mu} = {\dot x^{\mu}}$, with $u^{\mu}u_{\mu}=1$ defining its normalization. 

We consider a congruence of (non-intersecting geodesics) on our parameter manifold. 
If $\xi^{\mu}$ is the deformation vector joining two points on geodesics that are close, then its 
variation with the affine parameter can be used to characterize the congruence. 
Defining $B^{\mu}_{\phantom{1}\nu}=\nabla_{\nu}u^{\mu}$, we then have that  
$\nabla_{\nu}\xi^{\mu}u^{\nu} = B^{\mu}_{\phantom{1}\nu}\xi^{\nu}$ i.e $B^{\mu}_{\phantom{1}\nu}$
measures the amount by which $\xi^{\mu}$ fails to get parallel transported along the congruence. 
Here, $\nabla_{\mu}$ is defined by $\nabla_{\mu}V^{\nu} = \partial_{\mu}V^{\nu} + \Gamma^{\nu}_{\mu\lambda}V^{\lambda}$.
One can then write 
\begin{equation}
B_{\mu\nu}=\theta h_{\mu\nu}+\sigma_{\mu\nu}+\omega_{\mu\nu}
\label{evolution}
\end{equation}
where we have defined the projection tensor
$h_{\mu\nu}=g_{\mu\nu}-u_{\mu}u_{\nu}$. 
The expansion, shear and rotation parameters $\theta$, $\sigma^2$ and $\omega^2$ are then obtained from
\begin{eqnarray}
&~&\theta = B^{\mu}_{\phantom{1}\mu},~~~\sigma_{\mu\nu}  = \frac{1}{2}\left(B_{\mu\nu}+B_{\nu\mu}\right)-\theta h_{\mu\nu}~,\nonumber\\
&~&\omega_{\mu\nu} = \frac{1}{2}\left(B_{\mu\nu}-B_{\nu\mu}\right).
\label{esr}
\end{eqnarray}
In our case, these are scalar parameters that quantify changes in the shape and size of the 
congruence. We are interested in two dimensional parameter manifolds for which $\sigma_{\mu\nu}=\omega_{\mu\nu}=0$, thus
leaving $\theta$ as the only parameter that characterizes the congruence.

We use the same choice of parameters as in the last section, and with this, it can be checked from eq.(\ref{line}) that 
if we expand the metric around the critical point,
and write $T = T_c + t$, $\rho = \rho_c + r$, then the metric components up to first order are 
\begin{equation}
g_{TT} = \frac{3}{2}~,~~g_{\rho\rho} = t\left(1-r\right) \equiv t~,
\end{equation}
where we have neglected a cross term involving a product of $t$ and $r$. The scalar curvature near criticality is 
computed to be 
\begin{equation}
R = -\frac{1}{3t^2}
\label{nearc}
\end{equation}
This result agrees with that obtained in \cite{tapo4} for the van der Waals fluid. The third condition of the 
previous section is also satisfied. 

Importantly, we find that near criticality, the coordinate $r$ is cyclic. Writing the Lagrangian
\begin{equation}
{\mathcal L} = \frac{1}{2}\left(\frac{3}{2}{\dot t}^2 + t{\dot r}^2\right)~,
\end{equation}
the Euler-Lagrane equation along with the normalization condition 
can be used to compute $u^{\mu} = \left({\dot t}, {\dot r}\right)$ with the result
\begin{equation}
u^{\mu} = -\left[\left(\frac{2}{3}-\frac{2k^2}{3t}\right)^{1/2}, \frac{k}{t}\right]
\end{equation}
In conjunction with the discussion above, this is seen to yield
\begin{equation}
\theta = -\frac{1}{t\left(6-\frac{6k^2}{t}\right)^{1/2}}
\end{equation}
Now if we set $\lambda = 0$ at criticality, i.e measure the affine parameter from the critical point, then 
it is seen that we have to set $k=0$ so that $t=r=0$ at $\lambda = 0$. A non-zero value of
$k$ is also problematic as it makes the solution of ${\dot t}$ as well as $\theta$ imaginary near criticality.
With this condition, we find that $\lambda = -\sqrt{3/2}t$, so that 
\begin{equation}
\theta = \frac{1}{2\lambda}~,~~R = -\frac{1}{2\lambda^2}~,
\end{equation}
so that $R \sim \theta^2$. This result is strictly valid close to criticality, and is an example of a geometric scaling. 
In \cite{tapo4}, it was argued from generic grounds that this relation 
should in general be true for two dimensional parameter manifolds, and we see that it is 
indeed the case here. 

Here, we have chosen $c_v = 3/2$. In fact, an entirely similar computation can be performed with 
any value of $c_v$, in particular with $c_v\to 0$. We thus expect that results similar to those presented
in this section will be valid for RN-AdS black holes. 

\section{Discussions}
\label{sec6}

In this paper, we have considered the parameter space 
geometry of an RN-AdS fluid, which has a similar equation of state as that of the 
RN-AdS black hole. Such a fluid can be thought of as arising out of a virial expansion with
a non-zero fifth virial coefficient. Our equation of state for such a fluid is phenomenological in nature, and
depends on two constants. It reduces to that of the RN-AdS black hole as a special case, when a specific
identificaiton is made between these constants and the charge parameter of the black hole. 
In our construction, the RN-AdS fluid generically has a non-zero value of $c_v$ (which
we have exemplified by $c_v=3/2$). However, $c_v$ vanishes for the RN-AdS black hole, and we have shown that
the RN-AdS fluid, in the limit $c_v\to 0$ gives the same scalar curvature per unit volume, as that of the black hole. It thus seems
that the geometry of the RN-AdS black hole is more appropriately compared to that of the RN-AdS fluid in
the limit of vanishing $c_v$, rather than the vdW system. 

We have considered a line element which results in the curvature having an appropriate volume dimension. Here, we have
taken a fixed volume $V_s$, with the temperature $T$ and $\rho = N/V_s$ being the fluctuating variables, where $N$ denotes the 
particle number that is allowed to fluctuate, with a fixed $V_s$. 
We find that the scalar curvature has all the required properties, i.e 
has the appropriate dimension of volume, diverges along the spinodal curve and at criticality,  
it diverges with a critical exponent which equals $2$. 

These are known facts about the geometry of thermodynamics
of classical systems, and our method passes all the tests. We have numerically constructed geodesics
on the parameter manifold, and shown that they have qualitatively similar behavior as those in the vdW systems,
namely that they turn around near criticality and typically do not cross the spinodal curve. We have
further constructed a geometric scaling relation that bears strong resemblance to the vdW systems. We
also constructed the Widom lines for the RN-AdS fluid, both for a generic value of $c_v$, and for $c_v\to 0$,
and shown how these differ from the locus of maxima of $c_p$, slightly away from criticality. 

Interestingly, the RN-AdS fluid admits a minimum volume, and it is physically reasonable to demand that
the scalar curvature cannot be less than this volume, for the theory to be justifiable. This naturally implies an upper
cut-off for the pressure in the supercritical phase, beyond which the Gaussian approximation used
to derive the line element breaks down. This was noticed for the vdW fluid a while back, and a similar
feature persists in this case also. 

In this context, we note that in \cite{WeiLiuMannPRL}, the authors consider a similar system, but 
use the temperature $T$ and the volume $V_s$ as the variables for the metric, that are 
allowed to fluctuate. This is somewhat counter-intuitive, as in
this approach, the scalar curvature is dimensionless, which makes it difficult to identify it with a physical 
volume, thus reducing the scope of the analysis. Further, these authors
obtain $Rc_vt^2 = -1/8$ for the vdW fluid, which is different from $Rc_vt^2 = -1/2$ obtained from eq.(\ref{nearc}). This
last relation was also obtained long back by Ruppeiner (see eq.(6.62) of \cite{Rupp}). 

In this paper, we have treated the RN-AdS fluid phenomenologically. 
It will be interesting to understand the nature of these via statistical mechanics, and we hope to report on
this in the near future. It will also be interesting to explore further the physical properties of the RN-AdS fluid constructed here. 
Whether these can be mapped to real fluid systems is not clear at the moment, and we leave this for a future study. 

\begin{center}
{\bf Acknowledgements}
\end{center}
The work of T. S. is supported in part by Science and Engineering
Research Board (India) via Project No. EMR/2016/008037.


\begin{thebibliography}{}
\bibitem{Ray} 
  D.~Kastor, S.~Ray and J.~Traschen,
  %``Enthalpy and the Mechanics of AdS Black Holes,''
  Class.\ Quant.\ Grav.\  {\bf 26}, 195011 (2009).

\bibitem{Chamblin1}
A.~Chamblin, R.~Emparan, C.~V.~Johnson and R.~C.~Myers,
  %``Charged AdS black holes and catastrophic holography,''
  Phys.\ Rev.\ D {\bf 60}, 064018 (1999).

\bibitem{Chamblin2}
A.~Chamblin, R.~Emparan, C.~V.~Johnson and R.~C.~Myers,
  %``Holography, thermodynamics and fluctuations of charged AdS black holes,''
  Phys.\ Rev.\ D {\bf 60}, 104026 (1999).

\bibitem{KubiznakMann} 
  D.~Kubiznak and R.~B.~Mann,
  %``P-V criticality of charged AdS black holes,''
  JHEP {\bf 1207}, 033 (2012).

\bibitem{LL}
L. D. Landau and E. M. Lifshitz, {\tt Course of Theoretical Physics Vol 5 : Statistical Physics}, Pergamon
Press, 3rd Ed, 1980. 

\bibitem{Rupp}
G. Ruppeiner, Rev. Mod. Phys. {\bf 67}, 605 (1995), erratum {\it ibid} {\bf 68}, 313 (1996). 

\bibitem{PV}
J. P. Provost and G. Vallee, Comm. Math. Phys. {\bf 76} (1980) 289.

\bibitem{tapo5}
R.~Maity, S.~Mahapatra and T.~Sarkar,
  %``Information Geometry and the Renormalization Group,''
  Phys.\ Rev.\ E {\bf 92}, no. 5, 052101 (2015).

\bibitem{Widom}
B. Widom, Physica {\bf 73}, 107 (1974).

\bibitem{Widom1}
P. F. McMillan, H. E. Stanley, Nature Physics {\bf 6}, 479 (2010).

\bibitem{tapo1}
G.~Ruppeiner, A.~Sahay, T.~Sarkar and G.~Sengupta, Phys.\ Rev.\ E {\bf 86}, 052103 (2012).

\bibitem{NIST}
NIST Chemistry WebBook, available at the Web Site http://webbook.nist.gov/chemistry/.

\bibitem{tapo1b}
A.~Sahay, T.~Sarkar and G.~Sengupta,
  %``On the Thermodynamic Geometry and Critical Phenomena of AdS Black Holes,''
  JHEP {\bf 1007}, 082 (2010).

\bibitem{Sengupta}
P.~Chaturvedi, S.~Mondal and G.~Sengupta,
  %``Thermodynamic Geometry of Black Holes in the Canonical Ensemble,''
  Phys.\ Rev.\ D {\bf 98}, no. 8, 086016 (2018).

\bibitem{WeiLiuMannPRL}
S.~W.~Wei, Y.~X.~Liu and R.~B.~Mann,
  %``Repulsive Interactions and Universal Properties of Charged Anti–de Sitter Black Hole Microstructures,''
  Phys.\ Rev.\ Lett.\  {\bf 123}, no. 7, 071103 (2019).

\bibitem{RuppAJP}
G.~Ruppeiner, Am. J. Phys. {\bf 78}, 1170 (2010).

\bibitem{tapo6}
P.~Kumar, S.~Mahapatra, P.~Phukon and T.~Sarkar,
  %``Geodesics in Information Geometry : Classical and Quantum Phase Transitions,''
  Phys.\ Rev.\ E {\bf 86}, 051117 (2012).

\bibitem{tapo4}
P.~Kumar and T.~Sarkar,
  %``Geometric Critical Exponents in Classical and Quantum Phase Transitions,''
  Phys.\ Rev.\ E {\bf 90}, no. 4, 042145 (2014).

\end{thebibliography}
\end{document}